# Kinetics of Exciton Self-Trapping Induced Defect Accumulation in Rare-Gas Solids

A.N. Ogurtsov, O.N. Bliznjuk, N.Yu. Masalitina

*National Technical University "KhPI", Frunse Street 21, Kharkov 61002, Ukraine*

**Abstract**

The kinetics of the process of defect accumulation in rare-gas solids as a result of exciton self-trapping was studied using the selective vacuum ultraviolet photoluminescence method for monitoring of the crystal structure of the samples. The simple kinetic model of defect accumulation in rare-gas samples was applied to the fitting of the dose dependences of luminescence from solid Xe and Ne. The characteristic kinetic parameters were obtained by linear transformation of the time dependence of luminescence intensity of "defect" subbands.

## 1. Introduction

Rare-gas solids, or atomic cryocrystals, are widely used as model systems in fundamental investigations in physics and chemistry of solids [1-4], and as functional materials for particle detectors and most efficient positron moderators [5,6]. The high luminescence efficiency also underlies the proposed used of rare-gas solids for laser in the vacuum ultraviolet (VUV) [7], rare-gas matrices are widely used as media for unusual photoreactions [8], and thin films of rare-gas solids show the pronounced quantum-size effects [9]. As a consequence of the closed electronic shells, solid Xe, Kr, Ar, and Ne are the simplest known solids of the smallest binding energy, $\varepsilon_b$, between atoms in the lattice. On the other hand, solid Ar and Ne have band-gap energies, $E_g$, exceeding that of LiF and may be cited as the widest band-gap insulators. At the same time it is generally recognized that creation of electronic excitations in rare-gas solids results in self-destruction of crystal lattice. The investigation of the physics and chemistry of radiation-induced elementary processes of conversion of electronic excitations into structural defects in rare-gas solids is of considerable interest from the point of view of material modification by electronic excitations [10].

Contrary to conventional chemical or material engineering, which is mostly applied high temperature and/or high-pressures processes for material synthesis/modification and quite often the catalysts to speed up the reaction, the radiation is the unique source of energy, which can initiate chemical reactions at any temperature, under any pressure, in any phase, without use of catalysts [11]. The subthreshold inelastic radiation-induced atomic processes in rare-gas solids such as defect formation and desorption under excitation by photons and electrons with a kinetic energies below the threshold of knock-on of atoms from the lattice sites were studied recently [12-17]. Our study revealed that these processes have general similarity, they are all pass through the stage of trapping or self-trapping of mobile electronic excitations and, in principle, may be considered within the context of a single kinetic model [18,19], and that the harnessing of the intrinsic luminescence of rare-gas solids allows to carry out the real-time monitoring of the crystal structure of the samples [20]. However, despite the fact that a lot of information about electronic excitations in rare-gas solids has been documented in several books and reviews [12,21-24], up to now the kinetics and dynamics of elementary radiation-induced atomic processes in rare-gas solids were studied mainly qualitatively.

In the present paper we apply the simple kinetic model for excitonically-induced defect accumulation processes as a result of selective photoexcitation of rare-gas solids by synchrotron radiation, and propose the characteristic constants, which may be used for rare-gas samples certification and comparison.

## 2. Experimental

The experiments were carried out at the SUPERLUMI-station at HASYLAB, DESY, Hamburg. The selective photon excitation was performed by near-normal incidence 2 m primary monochromator with photon flux about 1012 photons/s at spectral resolution $\Delta\lambda = 0.2$ nm. The VUV-luminescence analysis was performed both with low-resolution, $\Delta\lambda = 2$ nm, Pouey high-flux monochromator equipped with a multisphere plate detector and with high-resolution, $\Delta\lambda = 0.1$ nm, secondary 1 m near-normal incidence monochromator equipped with a position-sensitive detector. The luminescence spectra were measured under steady state irradiation conditions, using a single-photon counting mode with accumulation time t = 1 s in every point of measurement, which allows to exclude the influence of pulsed nature of synchrotron radiation (with repetition frequency about 5 MHz) on progress curves of defect accumulation. The samples were grown in a special closed cell mounted on a He-flow cryostat holder in ultrahigh vacuum (UHV) environment (10–10 mbar). The experimental setup and methods of sample preparation from vapor phase were described in detail elsewhere [25,26]. In the present study we used the fast isobaric method of sample growing. The samples had diameter 1 cm and thickness 1 mm. The irradiated area of the sample was 4×0.15 mm$^2$ and to verify the reproducibility of the dose curves the several sequential independent irradiations of the same sample were performed by using of a sample movement mechanism.

## 3. Results and discussion

The electronic excitations in rare-gas solids have been under investigation since seventies and now the overall picture of creation and trapping of electronic excitations is basically complete. Because of strong interaction with phonons the excitons and holes in rare-gas solids are self-trapped, and a wide range of electronic excitations are created in samples: free excitons (FE), atomic-like (A-STE) and molecular-like self-trapped excitons (M-STE), molecular-like self-trapped holes (STH) and electrons trapped at lattice imperfections. Free and trapped excitations are coexist in rare-gas solids, which results in the presence of a wide range of luminescence bands with well-studied internal structure in their emission spectra (Fig. 1).

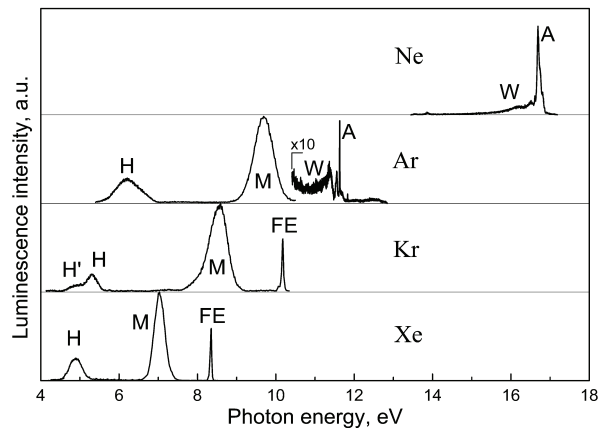

Fig. 1. Photoluminescence spectra of rare-gas solids at T=5 K under excitation by photons with energies $h\nu=E_g$.

Mutual interplay and evolution in time of the intensity and shape of different luminescence bands reveal the transformation of crystal properties of rare-gas samples under external effects [12,21-23]. On the other hand, the high sensitivity of luminescence spectra to sample growing and excitation conditions pose a real problem for comparison of the data from different crystals [15-17].

In the row Xe, Kr, Ar, Ne the atomic plarizability decreases as well as the crystal electron affinity; the latter becomes negative in the case of solid Ar and Ne [21,24]. As a consequence, the main channel of exciton self-trapping in Xe and Kr is the formation of M-STE, which is equivalent to a molecular dimer $R_2^*$ imbedded into the host lattice ($R$=rare gas atom). In the opposite case of solid Ne the negative electron affinity results in strong repulsion of the Rydberg orbital of the lowest $2p^53s$ excitation with the closed shell of the nearest neighbour atoms, which are pushed outwards. Such local lattice rearrangement leads to formation of A-STE, which is equivalent to an excited atom $R^*$ into the cavity (or "bubble"). In the intermediate case of solid Ar both M-STE and A-STE coexist. Since the energy of electronic excitation is transferred into kinetic energy of atomic motion over a unit cell, formation of three-, two-, or one-dimensional defects is ruled out. Only the point radiation defects, viz., Frenkel pairs may emerge in the bulk of the crystal.

In luminescence of solid Xe, Kr and Ar the most prominent feature is the so-called M-band. This band is formed by $^{1,3}\Sigma_u^+ \rightarrow ^1\Sigma_g^+$ transitions in ($R_2^*$) excimer M-STE. The negative electron affinity is a moving force of "bubble" formation around A-STE in the bulk of crystal, and desorption of atoms and excimers from the surface of solid Ne and Ar [24]. Radiative "hot" transitions in desorbed excimers of Ar and Ne result in a W-band [22]. A-bands are emitted by A-STE. Recent study of charged centers in rare-gas solids reveals the nature of H-bands [13]. These bands correspond to the third continua in rare gas emission, which are formed by transitions $(R_2^+)^* \rightarrow (R_2^+)$ in molecular ions [28]. A tiny amount of impurity Xe atoms in solid Kr results in the formation of heteronuclear excited ions $(KrXe^+)^*$ and a corresponding $H'$-band [12]. Radiative decay of free excitons from the bottom of the lowest $\Gamma(3/2)$, $n=1$ excitonic band produces $FE$-lines in spectra of solid Xe, Kr and Ar [22,28].

The analysis of the luminescence spectra of rare-gas solids under different excitation conditions, excitation energies and crystal-growth conditions made it possible to elucidate the internal structure of M and A bands. Each of the M-bands of Xe, Kr and Ar can be well approximated by two components (Fig. 2, a–c): low energy subband $M_1$ and high energy one $M_2$ [14].

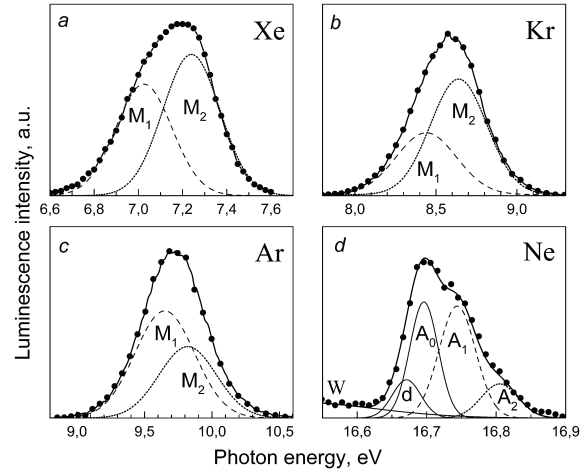

Fig. 2. Internal structure of luminescence bands under excitation by photons with energies: $h\nu_{Xe}$=9.15 eV, $h\nu_{Kr}$=11.21 eV, $h\nu_{Ar}$=13.57 eV, $h\nu_{Ne}$=20 eV.

The subband $M_2$ is dominant in the luminescence of more perfect samples. The spectra of samples with a great number of initial defects are mainly determined by the component $M_1$. This suggests that the subband $M_2$ is emitted by the excitons, which are self-trapped in the regular lattice, while the component $M_1$ is emitted by the centers, which are populated during trapping that occurs with the lattice imperfections involved. The excitation spectra of the $M_1$ and $M_2$ subbands were restored by decomposing the sequence of the luminescence spectra measured at different excitation energies [14]. Using the restored excitation spectra the threshold energies for the $M_1$ and $M_2$ subbands to appear were determined. The excitation spectra of the $M_1$-subbands exhibit preferential excitation at energies below the $n=1$ exciton and prove a direct photoabsorption by defect-related centers that produces the only $M_1$ component in the spectra [29]. In all cases the transformation of the $M$-band due to sample annealing or irradiation, resulting in the lattice degradation, may be described by the intensity redistribution between two subbands $M_1$ and $M_2$. The A-band of solid Ne associated with the transition $^3P_1 \rightarrow {}^1S_0$ consists of two main subbands: high-energy component $A_2$ stemmed from A-STE in a regular lattice and a low-energy one, $A_1$, which is associated with structural defects (Fig. 2, d). Additional subband $A_0$ stems from desorbed atoms. The $d$ component assigns to emission from metastable $^3P_2$ state. In previous papers [12,30] the subbands $d$, $A_0$, $A_1$, $A_2$ were marked as $d$, $b_0$, $b_1$, $b_2$. Similar structure has $A$-band of solid Ar [28].

The interatomic bond scission in the crystal lattice may be simulated either by elastic encounters between atoms composing solids and incoming particles or by creation of electronic excitations which transfer the energy to a specified crystal cell. The energy possessed by electronic excitation, no matter whether they are induced by the incidence of electrons, ions or photons, in a time range of an order of 0.1 ps via the processes of impact ionization, Auger transitions end electron-phonon interactions is relaxed into a large number of energy quanta correspondent to the band-gap energy or exciton energy [10].

The energy stored by such a low-energy electronic excitations in rare-gas solids is much higher than the binding energy $\varepsilon_b$ and various trapping processes concentrate the energy within a volume of about a unit cell. The extremely high quantum yield of luminescence [7] allows one to neglect non-radiative transitions, and the population of antibonding $^1\Sigma_g^+$ ground molecular state is usually considered as a main source of kinetic energy for a large-scale movement of atoms finishing in the Frenkel defects or desorption of atoms in the ground state – the ground-state mechanism [24]. On the other hand, the processes of formation of A-STE and M-STE centers themselves are accompanied by a considerable energy release to the crystal lattice, which also exceeds the binding energy $\varepsilon_b$ [12]. Such an excited-state mechanism of the large-scale atomic movement was studied recently [12,30,31].

The excited-state mechanism of M-STE to Frenkel pair conversion in solid Xe, Kr and Ar is supposed to occur by displacement of M-STE from centrosymmetric position in the <110> direction followed by reorientation to the <100> direction to stabilize the defect [31]. In the case of solid Ne the excited-state mechanism of A-STE to Frenkel pair conversion occurs by self-trapping of exciton accompanied by the strong repulsion of the Rydberg electron of excited atom with a closed shell of surrounding atoms and substantial local lattice rearrangement, which leads to a "bubble" formation around the excited atom [30]. After the bubble formation the surrounding ground state atoms are moved to the second shell and second-nearest neighboring vacancy-interstitial pairs could create the permanent defects, which remain in the lattice after exciton annihilation [33].

The time (or dose) dependences of the "defect" components "1" – the progress curves of the process of defect accumulation – turned out to be a very precise and sensitive tool to study the defect formation processes in rare-gas solids and provided the basis for the application of rare-gas solids in dosimetry [12]. Fig. 3 shows examples of such progress curves of solid Xe and Ne under irradiation by photons with energies $E < E_g$ [12,30].

An increase in the intensity of the defect component during irradiation reflects the accumulation of stable long-lived defects in the lattice as a result of exciton creation and self-trapping. Dose curves are saturated at long time of irradiation. Additional source of Frenkel pairs under irradiation by photons with energies $E > E_g$ is a hole self-trapping [24], which is also accompanied by creation of the metastable trapped centers [34] and shows a pronounced dose dependence of H-band [35].

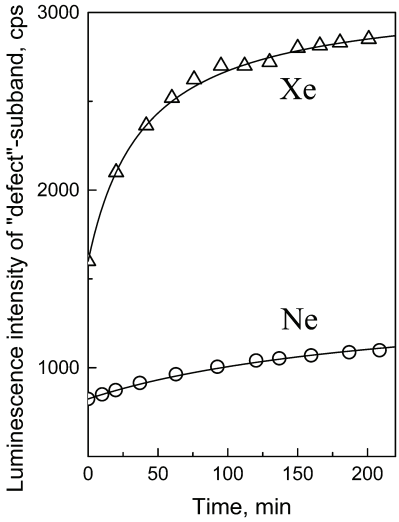

Fig. 3. Dose dependences of subband M1 of solid Xe excited by photons with $h\nu = 9.15$ eV ($\triangle$) and subband $A_1$ of solid Ne excited by photons with $h\nu = 20$ eV ($\circ$), and their fitting by Eq. (7) (solid curve).

In all cases we can consider the process of defect formation as a combination of three separate processes:

$$E + T \xrightarrow{k_1} MTE, \qquad (1)$$

$$MTE \xrightarrow{k_{-1}} T, \qquad (2)$$

$$MTE \xrightarrow{k_2} D. \qquad (3)$$

Process (1) is a trapping of mobile excitation, $E$, with a rate constant $k_1$, on trapping center, $T$, and formation of an excited metastable trapped center $MTE$ (A-STE, M-STE, or STH), which may be considered as metastable short-lived lattice defect. In the case of perfect lattice the self-trapping of mobile excitations can occur at every atomic position within the lattice. Radiative decay of the short-lived $MTE$-center either returns the lattice with a rate constant $k_{-1}$ into the initial state without permanent defect (process (2)), or forms the permanent defect $D$ (Frenkel pair) in the process (3) with a rate constant $k_2$.

The rate equations for concentrations of $MTE$-centers, trapping centers and permanent defects are

$$\frac{dn_{MTE}}{dt} = k_1 n_E n_T - k_{-1} n_{MTE} - k_2 n_{MTE}, \qquad (4)$$

$$\frac{dn_T}{dt} = k_{-1} n_{MTE} - k_1 n_E n_T, \qquad (5)$$

$$\frac{dn_D}{dt} = k_2 n_{MTE}. \qquad (6)$$

We assume that low level of irradiation under steady state conditions creates a constant low concentration of mobile excitations, $N_0$, which is much less then concentration of the trapping centers, $n_T \gg N_0$, and at the beginning of irradiation $n_T$ can be taken as constant. Consequently, from Eq. (5) we can determine the constant $A_1 = k_{-1}/k_1 = n_E n_T (n_{MTE})^{-1}$, and, taking into account the fact, that irradiation produced electronic excitations may be either free, or trapped, $N_0 = n_E + n_{MTE}$, one can express Eq. (6) as $dn_D/dt = A_2$, where $A_2 = k_2 N_0 n_T (A_1 + n_T)^{-1}$ is the constant. Thus, at the beginning of irradiation (concentration of defects $n_D$ is small), defect concentration indeed grows linear with time, $n_D = A_2 \cdot t$. Linear growth of "1"-subbands at the beginning of irradiation was experimentally detected in all experiments. As was shown earlier [25] the time dependence of luminescence intensity of "defect" subband, $I_1(t)$, under steady-state conditions may be expressed in form

$$I_1(t) = I_1(0) + \frac{K \cdot t}{L + t}, \qquad (7)$$

where $I_1(0)$ is the initial intensity of "defect" luminescence due to $n_D \neq 0$ at $t = 0$; $K = k_2 N_0$ is the saturation value of $(I_1(t) - I_1(0))$ at $t \to \infty$; $L = (k_{-1} + k_2)/(k_1 A_2)$ is a characteristic constant of a sample – under identical excitation and detection conditions the sample with less pronounced processes of defect formation will have a higher value of $L$. Fig. 3 shows the examples of fitting by Eq. (7) the dose curves of "defect" components "1" of solid Xe and Ne.

In the case when it is not possible to obtain the totally saturated progress curve, one can determine the values of the constants $K$ and $L$ from interceptions of the straight line of data linear fit in the double-reciprocal plot of $(I_1(t) - I_1(0))^{-1}$ versus $t^{-1}$. Linear transformation of Eq. (7) in the form

$$\frac{1}{I_1(t) - I_1(0)} = \frac{1}{K} + \frac{L}{K}\frac{1}{t} \qquad (8)$$

allows one to determine the values of constants $K$ and $L$ from the double reciprocal plot (Fig. 4) for the particular case of progress curves of defect accumulation, which are shown at Fig. 3.

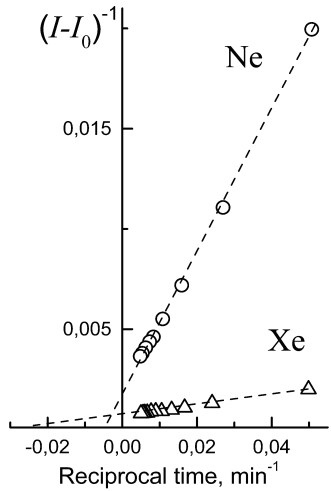

Fig. 4. Fitting of data from Fig. 3 in double reciprocal plot by Eq. (8).

The fitting values of constants are $K_{Xe}$ = 1600 cps, $K_{Ne}$ = 600 cps, $L_{Xe}$ = 2.4·10³ s, $L_{Ne}$ = 1.4·10⁴ s, which is in line with general increase of defect formation efficiency in the sequence Ne, Ar, Kr, Xe [24]. Analytical application of the model enable one to compare different rare-gas crystals with standard one and estimate the radiation doses from initial part of the dose curves.

## 4. Summary

The application of rare-gas solids as functional materials in high-energy physics and dosimetry inevitably faces fundamental problem of materials modification by electronic excitation. Even selective excitation of excitons in rare-gas solids by photons with energies $E < E_g$ results in accumulation of Frenkel-pairs by intrinsic excited-state and ground-state mechanisms of defect formation induced by self-trapping of excitons. The harnessing of rich luminescence spectra of rare-gas solids for real-time monitoring of their crystal structure and application of the simple kinetic model allow the fitting of the experimental progress curves in the form of dose dependencies of "defect" luminescence subbands and obtaining the particular kinetic parameters. This approach provides a way of qualitative and quantitative analysis and certification of rare-gas crystals, which is indispensable at any attempt of comparison of data from experiments with different samples.